\documentclass{article}
\usepackage{spconf,graphicx, amssymb}
\usepackage{subcaption}
\usepackage{bm}
\usepackage{multirow}
\usepackage{booktabs}
\usepackage{tikz}
\usepackage{amsmath,amsfonts}
\usepackage{pifont}
\newcommand{\xmark}{\ding{55}}%
\newcommand{\cmark}{\ding{51}}%

\title{All-neural beamformer for continuous speech separation}
\name{\begin{tabular}{c}
Zhuohuang Zhang$^{1,2}$, Takuya Yoshioka$^{1}$, Naoyuki Kanda$^{1}$, Zhuo Chen$^{1}$, Xiaofei Wang$^{1}$,\\
Dongmei Wang$^{1}$, Sefik Emre Eskimez$^{1}$ \thanks{This work was performed during an internship at Microsoft.}
\end{tabular}}

\address{$^1$Microsoft, Redmond, WA, USA~~~~~~~~~~
$^2$ Indiana University, Bloomington, IN, USA\\{\small \texttt{zhuozhan@iu.edu, \{tayoshio, nakanda, zhuc, xiaofewa, dowan, seeskime\}@microsoft.com}}}

\begin{document}
\ninept
\maketitle
\begin{abstract} 
Continuous speech separation (CSS) aims to separate overlapping voices from a continuous influx of conversational audio containing an unknown number of utterances spoken by an unknown number of speakers. A common application scenario is transcribing a meeting conversation recorded by a microphone array. Prior studies explored various deep learning models for time-frequency mask estimation, followed by a minimum variance distortionless response (MVDR) filter to improve the automatic speech recognition (ASR) accuracy. The performance of these methods is fundamentally upper-bounded by MVDR's spatial selectivity. Recently, the all deep learning MVDR (ADL-MVDR) model was proposed for neural beamforming and demonstrated superior performance in a target speech extraction task using pre-segmented input. In this paper, we further adapt ADL-MVDR to the CSS task with several enhancements to enable end-to-end neural beamforming. The proposed system achieves significant word error rate reduction over a baseline spectral masking system on the LibriCSS dataset. Moreover, the proposed neural beamformer is shown to be comparable to a state-of-the-art MVDR-based system in real meeting transcription tasks, including AMI, while showing potentials to further simplify the runtime implementation and reduce the system latency with frame-wise processing. 
\end{abstract}

\section{Introduction}
\label{sec:intro}
Undesirable background noises or interfering speakers often contaminate speech in daily communications. This poses a significant challenge for the current automatic speech recognition (ASR) systems as they are designed for the scenario where at most one person is speaking at a given time instance. Speech separation algorithms have been proposed to address this issue by separating different speaker sources from a mixture signal. Speech separation algorithms have been serving as important front-ends for various different speech communication systems, including ASR \cite{zhang2017speech,wang2018spatial}, meeting transcription \cite{yoshioka2019advances}, and digital hearing-aid devices \cite{doclo2010acoustic}. 

With the recent advancements in deep learning, several data-driven speech separation algorithms have been proposed \cite{yoshioka2018multi,luo2020end,chen2021continuous}, yielding improved speech quality and intelligibility. These systems include time-frequency (T-F) mask-based systems \cite{zhang2020loss,yu2020constrained}, and some other time-domain end-to-end systems such as TasNet \cite{luo2018tasnet}, Conv-TasNet \cite{luo2019conv} and Wave-U-Net \cite{stoller2018wave}. However, these purely deep learning-based systems focus on removing undesired interfering sources without having constraints for limiting the solution space. This often results in non-linear distortions on the separated speech that are harmful to the current ASR systems \cite{xu2021generalized}.

The minimum variance distortionless response (MVDR) filter has been widely adopted to address this non-linear distortion issue. It is often combined with a neural network that estimates the speech and noise components for deriving the beamforming filtering weights. However, the mathematically-derived MVDR solution is not straightforward for end-to-end optimization due to numerical instability \cite{zhang2021end}. 
For the same reason, it is also challenging to perform adaptive beamforming on a frame-by-frame basis stably. Therefore, the MVDR filter is usually applied on a per-segment basis. Recently, all deep learning MVDR (ADL-MVDR) \cite{zhang2021adl,zhang2020multi} was proposed for neural frame-adaptive beamforming and demonstrated superior performance in audio quality of the separated signals and the ASR accuracy in a target speech extraction task for pre-segmented speech mixtures. It incorporates two separate gated recurrent unit (GRU) \cite{cho2014properties} based networks to replace the matrix operations (e.g., matrix inverse) involved in the conventional MVDR solution, which bypasses the numerical instability issue and makes the end-to-end training more feasible. 

In this paper, we extend ADL-MVDR to continuous speech separation (CSS) to enable frame-wise neural beamforming. Unlike most of the existing studies that convert pre-segmented audio mixture into per-speaker separated speech, a CSS system converts long-form unsegmented audio, including an unknown number of speakers into a few (two in our experiments) audio streams, each of which contains overlap-free signals~\cite{ccetin2006analysis}. This CSS design enables us to handle overlapping speech of an unknown number of speakers with low latency, which is preferable to serve for many real applications such as meeting transcription.
We introduce several enhancements to make ADL-MVDR effective for the CSS setting, including steering vector normalization, the use of a voice activity detection (VAD) network, positive semi-definite constraint on matrix inversion and residual connection. 
The proposed neural beamformer is first evaluated on the LibriCSS \cite{chen2020continuous} dataset, which consists of long-form multi-talker real recordings generated by concatenating and mixing utterances from LibriSpeech dataset \cite{panayotov2015librispeech}. We further compare our systems' performance on several real meeting recordings (including AMI corpus \cite{carletta2005ami} and Microsoft internal meetings), where the proposed neural beamformer is shown to be comparable to a state-of-the-art MVDR-based system and demonstrates potentials on reducing system latency with frame-wise beamforming. 


\vspace{-2mm}
\section{Technical Background}
\label{sec:definition}

\subsection{Continuous Speech Separation}
\label{subsec:CSS}
\vspace{-.3em}

\begin{figure}[tb!]
  \centering
  \includegraphics[scale = 0.73]{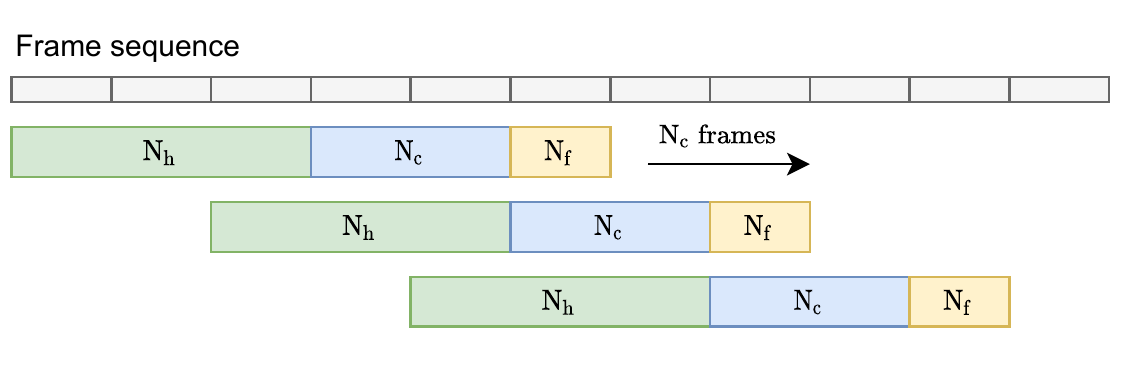}
  \vspace{-1em}
  \caption{The chunk-wise processing scheme for CSS is shown.}
  \label{fig:fig1}
  \vspace{-3mm}
\end{figure}

Most of the existing speech separation algorithms operate on pre-segmented mixtures by assuming an ideal overlap detector to be available. However, in real scenarios, speech separation systems need to deal with a continuous audio stream consisting of multiple speakers, which can be hours long and include both overlapped and non-overlapped utterances. More recently, the CSS scheme \cite{yoshioka2019low} has been proposed, which is defined as the process of generating a limited number of overlap-free signals from the continuous audio stream. To deal with the long input signals, we adopt the chunk-wise CSS scheme proposed in \cite{yoshioka2018recognizing}. As illustrated in Fig. \ref{fig:fig1}, a sliding window is used which contains three sub-windows, including the history sub-window ($N_h$ frames), current sub-window ($N_c$ frames), and future sub-window ($N_f$ frames). For each time step, the window is moved forward by $N_c$ frames. During test time evaluation, the speech separation algorithm takes in the entire chunk of information (i.e., $N = N_h + N_c + N_f$ frames) to estimate $K (=2)$ overlap-free signals for the current $N_c$ frames. Estimated signals for each chunk are later aligned via block stitching \cite{chen2020continuous}. 

\subsection{ADL-MVDR for Target Speech Separation}
\label{subsec:ADL-MVDR}
\vspace{-.3em}

The MVDR filter aims to preserve the information from the target direction while minimizing the power of interfering sources. Specifically, the MVDR filter is defined as
\begin{equation}
\label{eq:mvdr-constraint}
\mathbf{h_{\text{MVDR}}}=\underset{\mathbf{h}}{\arg \min \mathbf{h}}^{\mathrm{H}} \mathbf{\Phi}_{\mathrm{VV}} \mathbf{h} \quad \bf{\text {s.t.}} \quad \mathbf{h}^{\mathrm{H}} \mathbf{\boldsymbol{v}}=\mathbf{1}, 
\end{equation}
where $\mathbf{h_{\text{MVDR}}}$ is the MVDR filtering weights, superscript $^{\mathrm{H}}$ denotes the Hermitian transpose, and $\mathbf{\Phi}_{\mathrm{VV}}$ is the covariance matrix of interfering sources (background noise and/or interfering speakers). Variable $\mathbf{\boldsymbol{v}}$ represents the steering vector which can be approximated by extracting the principal eigenvector of the speech covariance matrix \cite{liu2018neural}, i.e., $\mathbf{\boldsymbol{v}}=\mathcal{P}\{\mathbf{\Phi}_{\mathrm{SS}}\}$. By solving Eq. (\ref{eq:mvdr-constraint}), we have \cite{higuchi2017online,shimada2018unsupervised}
\begin{equation}
\label{eq:mvdr-solution}
\mathbf{h_{\text{MVDR}}}=\frac{\mathbf{\Phi}_{\mathrm{VV}}^{-1} \boldsymbol{v}}{\mathbf{\boldsymbol{v}^{\mathrm{H}}} \mathbf{\Phi}_{\mathrm{VV}}^{-1} \mathbf{\boldsymbol{v}}}. 
\end{equation}
Most existing studies estimate the covariance matrices in a chunk-wise fashion. However, this chunk-wise processing also incurs less flexibility and adaptability of the MVDR filter. 

ADL-MVDR has been recently proposed as a fully neural network-based beamformer that has demonstrated superior performance in a target speech extraction task \cite{zhang2021adl,zhang2020multi}. The core idea of ADL-MVDR is utilizing two separate GRU-based networks (denoted as GRU-Nets) to replace the matrix inversion and principal eigenvector extraction involved in the conventional MVDR solution. Each of the GRU-Nets takes in the speech/noise covariance matrix and estimates the steering vector or the matrix inverse on a per-frame basis. Note that, unlike the conventional MVDR systems, which take sums or expectations over the entire chunk to derive the covariance matrices, the temporal information in the covariance matrices can be leveraged in the ADL-MVDR network. 

\vspace{-2mm}
\section{Proposed System}
\label{sec:proposedsystem}

\begin{figure*}[t!]
  \centering
  \includegraphics[scale = 0.76]{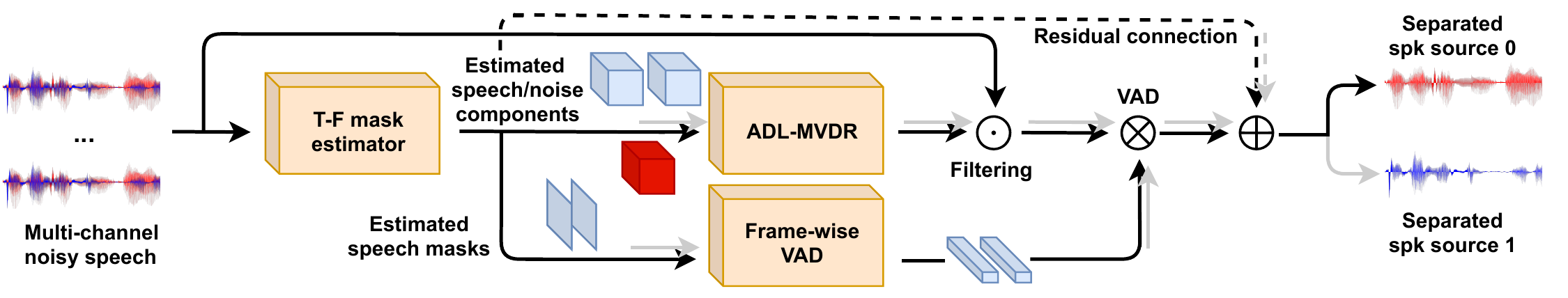}
  \caption{Diagram of proposed neural beamformer for continuous speech separation is shown. $\bigodot$ and $\bigotimes$ represent the operations described in Eqs. (\ref{eq:tf-beamforming}) and (\ref{eq:vad}), respectively. $\bigoplus$ denotes the addition. The gray arrow indicates the flow pass of the second source.}
  \label{fig:overview}
  \vspace{-3mm}
\end{figure*}

The overall framework of our proposed system is depicted in Fig. \ref{fig:overview}. The T-F mask estimator first predicts three T-F masks for CSS tasks, including two-speaker masks and an isotropic noise mask. ADL-MVDR then estimates time-varying beamforming weights based on the estimated masks. Meanwhile, we introduce a VAD network based on the GRU-Net, whose result is multiplied with the output of ADL-MVDR. Finally, the output audio is emitted with an additional residual connection from the T-F masked speech signal. During training, the entire system is updated with permutation invariant training (PIT) scheme \cite{yu2017permutation}. 

\subsection{ADL-MVDR for CSS}
\label{subsec:ADL-MVDR}
\vspace{-.3em}

In the CSS scheme, at most $K$ speech sources are assumed to be active within each chunk. Here, we describe the proposed method with $K=2$. The normalized time-varying input covariance matrices for the speaker and interfering noise sources can be derived as
\begin{equation}
\begin{aligned}
\label{eq:framewise_cov}
\mathbf{\Phi}_{\mathrm{SS}}^{(k)}(t,f) &= \frac{\mathbf{\hat{S}_{\text{mask}}}^{(k)}(t,f) \mathbf{\hat{S}_{\text{mask}}}^{\mathrm{H}\:(k)}(t,f)}{\sum_{t=1}^{T} \mathbf{M}^{(k)2}_{\mathrm{S}}(t,f)}, \\
\mathbf{\Phi}_{\mathrm{VV}}^{(k)}(t,f) &= \frac{\mathbf{\hat{V}_{\text{mask}}}^{(k)}(t,f) \mathbf{\hat{V}_{\text{mask}}}^{\mathrm{H}\:(k)}(t,f)}{\sum_{t=1}^{T} \mathbf{M}^{(k)2}_{\mathrm{V}}(t,f)},
\end{aligned}
\end{equation}
where $\mathbf{\Phi}_{\mathrm{SS}}^{(k)}$ is the instantaneous speech covariance matrix for speaker source $k\in\{0,1\}$, and $(t,f)$ represent the time and frequency indices. $\mathbf{\hat{S}_{\text{mask}}}^{(k)} = \mathbf{M}^{(k)}_{\mathrm{S}} \mathbf{Y}$ is the masked speech for source $k$, where $\mathbf{Y}$ is the multi-channel noisy speech and $\mathbf{M}_{\mathrm{S}}^{(k)}$ denotes the T-F speech mask for source $k$. $T$ is the total number of frames. 
Similarly,  $\mathbf{\Phi}_{\mathrm{VV}}^{(k)}$ is the interfering source covariance matrix for source $k$. The interfering source consists of two parts (i.e., ambient noise and interfering speaker) as $\mathbf{\hat{V}_{\text{mask}}}^{(k)}= \mathbf{M}_{\mathrm{N}} \mathbf{Y}+\mathbf{\hat{S}_{\text{mask}}}^{(1-k)}$, where $\mathbf{M}_{\mathrm{N}}$ is the isotropic noise mask. Similarly we have $\mathbf{M}_{\mathrm{V}}^{(k)} = \mathbf{M}_{\mathrm{N}} + \mathbf{M}_{\mathrm{S}}^{(1-k)}$. Next, the time-varying variables corresponding to the steering vector and inverse noise covariance matrix are estimated using two separate GRU-Nets as
\begin{equation}
\begin{aligned}
\mathbf{\boldsymbol{\hat{\boldsymbol{v}}}}^{(k)}(t,f) & = \mathbf{GRU{\text -}Net}_{\upsilon}(\mathbf{\Phi}_{\mathrm{SS}}^{(k)}(t,f)), \\
{\mathbf{\hat{\Phi}}_{\mathrm{VV}}}^{-1\:(k)}(t,f) & = \mathbf{GRU{\text -}Net}_{\mathrm{VV}}(\mathbf{\Phi}_{\mathrm{VV}}^{(k)}(t,f)).
\end{aligned}
\end{equation}
Once these time-varying coefficients are obtained, the beamforming weights $\mathbf{h}_{\text{ADL-MVDR}} \in \mathbb{C}^{F\times T \times C}$ ($F$ and $C$ represent the frequency and channel dimensions, respectively) can be derived on a per-frame basis by plugging these estimated terms into Eq. (\ref{eq:mvdr-solution}). Finally, the ADL-MVDR filtered speech for source $k$ can be obtained as
\begin{equation}
\label{eq:tf-beamforming}
\mathbf{\hat{S}}^{(k)}_{\text{ADL-MVDR}}(t,f) = \mathbf{\hat{h}}^{\mathrm{H}\:(k)}_{\text{ADL-MVDR}}(t,f)\mathbf{Y}(t,f).
\end{equation}

\begin{figure}[t!]
  \centering
  \includegraphics[scale = 0.42]{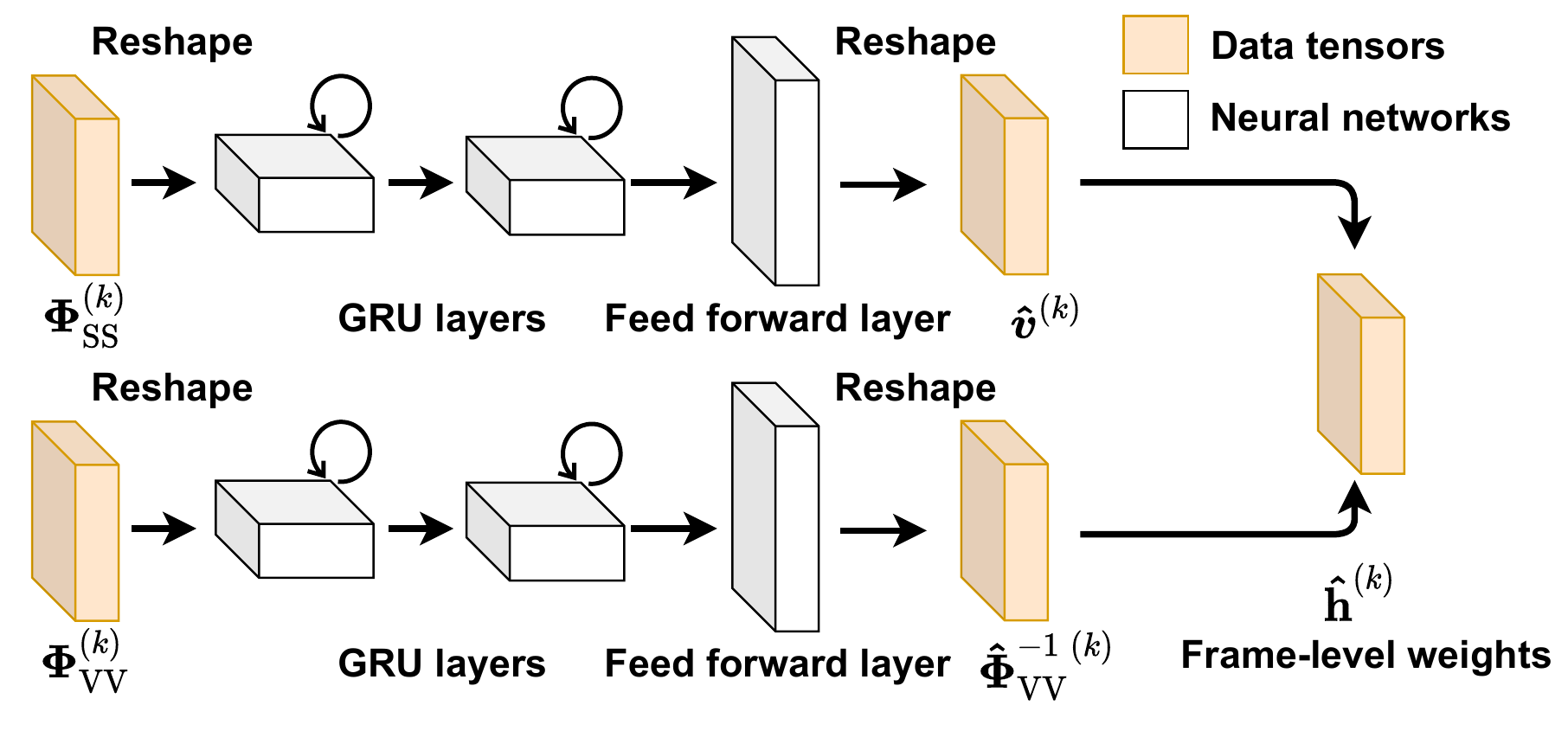}
  \caption{ADL-MVDR module is shown. It consists of two separate GRU-Nets to replace the principal eigenvector extraction and matrix inversion, respectively. The estimated frame-level variables are then used to derive the frame-level beamforming weights.}
  \label{fig:adlmvdrcss}
  \vspace{-3mm}
\end{figure}

\subsection{Enhancements to ADL-MVDR}
\label{subsec:enhanced_ADL-MVDR}
\vspace{-.3em}

To further enhance the performance of the ADL-MVDR in the CSS scheme, we propose several techniques. Firstly, we introduce constraints on the estimated steering vector and the inverse of the covariance matrix. Specifically, we apply normalization on the estimated time-varying steering vector, where the normalized steering vector is derived as $\bar{\boldsymbol{v}}=\hat{\boldsymbol{v}}/|\hat{\boldsymbol{v}}|$. A positive semi-definite constraint is also imposed on the estimated inverse of the interfering noise covariance matrix. This is done by modifying the GRU-Net to estimate an upper triangular matrix $U$, by which the inverse matrix is calculated as $\mathbf{\hat{\Phi}}^{-1}_{\mathrm{VV}} = UU^H$.

Secondly, as we mentioned at the beginning of this section, we introduce the VAD network to control the gain of the output. 
\begin{equation}
\begin{aligned}
\label{eq:vad}
\hat{\mathbf{w}}_{\text{VAD}}^{(k)}(t) &= \mathbf{GRU{\text -}Net}_{\text{VAD}}(\mathbf{M}_{\mathrm{S}}^{(k)}(t,f)), \\
\hat{\mathbf{S}}_{\text{VAD}}^{(k)}(t,f) &= \hat{\mathbf{S}}_{\text{ADL-MVDR}}^{(k)}(t,f)\hat{\mathbf{w}}_{\text{VAD}}^{(k)}(t),
\end{aligned}
\end{equation}
where $\hat{\mathbf{S}}_{\text{VAD}}^{(k)}$ is the VAD filtered speech for source $k$, $\hat{\mathbf{w}}_{\text{VAD}}^{(k)} \in \mathbb{R}$ denotes the estimated frame-level VAD weights. Finally, the separated speech $\hat{\mathbf{S}}^{(k)}$ can be obtained with residual connection as
\begin{equation}
\label{eq:residual}
\hat{\mathbf{S}}^{(k)}(t,f) = \hat{\mathbf{S}}_{\text{VAD}}^{(k)}(t,f) + \alpha \cdot \mathbf{\hat{S}_{\text{mask}}}^{(k)}(t,f),
\end{equation}
where $\alpha$ is a weighting factor, and we use the first channel of the masked speech for the residual connection.

\subsection{Mask Estimator and GRU-Nets}
\label{subsec:baseline}
\vspace{-.3em}

We use a T-F mask estimation model proposed in \cite{yoshioka2021vararray}. The multi-channel input speech stream is first encoded by a shared encoding block (based on conformer layers \cite{gulati20_interspeech,chen2021continuous}) to extract the intra-channel feature independently, followed by a stack of geometry agnostic modules. Within each geometry agnostic module, there is a transform-average-concatenate (TAC) block \cite{luo2020end} and another shared conformer to alternately encode the inter- and intra-channel information. Finally, the encoded features are pooled out by taking the average across the channel dimension and fed into another conformer block to estimate the three T-F masks (i.e., two speakers and one noise~\cite{yoshioka2018recognizing}). 

Each GRU-Net consists of two unidirectional GRU layers, followed by another feed forward layer (FFL) as illustrated in Fig.~\ref{fig:adlmvdrcss}. For frame-wise coefficients estimation, the GRU-Net takes in the covariance matrix that is derived following Eq. (\ref{eq:framewise_cov}). Note that the real and imaginary parts are concatenated as input. We use the same architecture for both the ADL-MVDR and VAD networks except the difference in the number of units in each layer.



\section{Experimental Setup}
\label{sec:experimentalsetup}

\subsection{Datasets}
\label{subsec:dataset}
\vspace{-.3em}

Our training set contains 219 hours of randomly mixed and reverberated utterances from WSJ1 SI-284~\cite{linguistic1994csr}. We simulated the multi-channel mixtures by randomly picking the audio of one or two speakers and convolving it with a 7-channel room impulse response, which was simulated with the image method~\cite{allen1979image}. Then, the reverberated signals were mixed with a source energy ratio between -5 and 5 dB. Simulated isotropic noise was then added with a 0-10 dB signal-to-noise ratio (SNR). Noise samples from MUSAN~\cite{musan2015} were also reverberated and added at an SNR between -5 and 10 dB. The average overlap ratio of the training set was about 50\%.

LibriCSS \cite{chen2020continuous} was used for the first experiment, which consists of 10 hours of 7-channel recordings. Sound sources were generated by mixing LibriSpeech utterances \cite{panayotov2015librispeech}. They were played back in a real meeting room and recorded by a 7-channel microphone array. Several overlap ratios were included, ranging from 0 to 40\%. There were two subsets for the 0\% overlap condition: one with short (S) inter-utterance silence and one with long (L) inter-utterance silence.

To evaluate the performance of the proposed neural beamformer in more realistic and challenging environments, we also carried out experiments using real meeting datasets; namely, the AMI corpus \cite{carletta2005ami} and Microsoft internal meeting dataset, dubbed as MS. The AMI recordings were made with an 8-channel circular microphone array, while the MS data collection used a 7-channel array used in \cite{chen2020continuous}.
MS contained 60 sessions in total with various numbers of speakers per session. 
In order to get the transcriptions, we adopted a modified version of the conversation transcription system described in \cite{yoshioka2019advances} with a hybrid ASR model.

\begin{table*}[htb!]
\centering
\caption{Continuous speech separation results on LibriCSS dataset are shown. The numbers in the left and right of `/' symbol indicate the word error rates (WERs) with real- and complex-valued masks, respectively. Best performance is marked with \vspace{-.7em}
\textbf{bold} font for each condition.}
\label{tab:tab1}
\scalebox{0.81}{
\begin{tabular}{cccccccccccccc}
\hline \hline
\multirow{2}{*}{\textbf{\begin{tabular}[c]{@{}c@{}}Sys.\\ ID\end{tabular}}} & \multirow{2}{*}{\textbf{\begin{tabular}[c]{@{}c@{}}Loss \\ type\end{tabular}}} & \multirow{2}{*}{\textbf{Beamformer}} & \multirow{2}{*}{\textbf{\begin{tabular}[c]{@{}c@{}}Norm. \\ $\boldsymbol{v}$\end{tabular}}} & \multirow{2}{*}{\textbf{\begin{tabular}[c]{@{}c@{}} Positive semi-\\ definite $\mathbf{\Phi}^{-1}_{\mathrm{VV}}$\end{tabular}}} & \multirow{2}{*}{\textbf{VAD}} & \multicolumn{1}{c|}{\multirow{2}{*}{\textbf{\begin{tabular}[c]{@{}c@{}}Res. \\ connection\end{tabular}}}} & \multicolumn{6}{c|}{\textbf{Overlap ratios (\%)}} & \textbf{} \\
 &  &  &  &  &  & \multicolumn{1}{c|}{} & \textbf{0L} & \textbf{0S} & \textbf{10} & \textbf{20} & \textbf{30} & \multicolumn{1}{c|}{\textbf{40}} & \textbf{Avg.} \\ \hline
\multicolumn{14}{c}{Baseline spectral masking systems} \\ \hline
0 & Mag. & N/A & N/A & N/A & N/A & \multicolumn{1}{c|}{N/A} & \textbf{6.1}/6.6 & \textbf{6.9}/7.3 & \textbf{8.6}/9.8 & \textbf{11.4}/12.0 & \textbf{14.7}/15.3 & \multicolumn{1}{c|}{\textbf{15.8}/16.2} & \textbf{11.1}/11.7 \\
1 & Log-mel & N/A & N/A & N/A & N/A & \multicolumn{1}{c|}{N/A} & 6.4/6.8 & \textbf{6.9}/7.2 & 9.2/9.6 & 11.9/12.1 & 15.1/15.1 & \multicolumn{1}{c|}{16.7/17.1} & 11.6/11.9 \\ \hline
\multicolumn{14}{c}{Proposed neural beamformer systems} \\ \hline
2 & Mag. & ADL-MVDR & \xmark & \xmark & \xmark & \multicolumn{1}{c|}{\xmark} & 9.1/9.4 & 9.6/8.9 & 11.5/11.3 & 13.5/14.4 & 16.7/16.3 & \multicolumn{1}{c|}{18.5/17.8} & 13.7/13.5 \\
3 & Mag. & ADL-MVDR & \xmark & \xmark & \cmark & \multicolumn{1}{c|}{\xmark} & 6.1/6.3 & 6.8/6.6 & 9.1/9.0 & 11.5/11.8 & 14.6/14.1 & \multicolumn{1}{c|}{16.1/15.7} & 11.3/11.1 \\
4 & Mag. & ADL-MVDR & \cmark & \xmark & \cmark & \multicolumn{1}{c|}{\xmark} & \textbf{5.9}/6.1 & 6.5/6.7 & 9.0/9.0 & 11.3/11.1 & 13.9/13.6 & \multicolumn{1}{c|}{15.4/15.0} & 10.9/10.7 \\
5 & Mag. & ADL-MVDR & \cmark & \xmark & \cmark & \multicolumn{1}{c|}{\cmark} & 6.1/6.1 & 6.7/\textbf{6.4} & \textbf{8.7}/9.3 & \textbf{10.5}/11.8 & \textbf{13.2}/14.2 & \multicolumn{1}{c|}{\textbf{14.7}/15.3} & \textbf{10.5}/11.0 \\
6 & Mag. & ADL-MVDR & \cmark & \cmark & \cmark & \multicolumn{1}{c|}{\cmark} & 6.3/6.1 & 6.5/\textbf{6.4} & 8.8/9.2 & 11.2/11.6 & 13.4/14.0 & \multicolumn{1}{c|}{15.1/15.4} & 10.7/11.0 \\ \hline
7 & Log-mel & ADL-MVDR & \xmark & \xmark & \cmark & \multicolumn{1}{c|}{\xmark} & 6.0/\textbf{5.8} & 6.5/6.4 & 9.3/8.6 & 11.5/10.8 & 14.1/13.2 & \multicolumn{1}{c|}{15.6/15.1} & 11.0/10.5 \\
8 & Log-mel & ADL-MVDR & \cmark & \xmark & \cmark & \multicolumn{1}{c|}{\xmark} & 6.0/6.4 & 6.3/6.4 & 8.7/\textbf{8.4} & 10.7/11.1 & 12.7/13.1 & \multicolumn{1}{c|}{14.5/15.0} & 10.3/10.5 \\
9 & Log-mel & ADL-MVDR & \cmark & \xmark & \cmark & \multicolumn{1}{c|}{\cmark} & 6.0/6.1 & 6.6/6.4 & 8.6/8.8 & 10.9/11.4 & 12.7/13.8 & \multicolumn{1}{c|}{\textbf{14.0}/15.1} & 10.2/10.8 \\
10 & Log-mel & ADL-MVDR & \cmark & \cmark & \cmark & \multicolumn{1}{c|}{\cmark} & 6.0/6.0 & 6.3/\textbf{6.0} & 8.5/8.6 & \textbf{10.5}/11.1 & \textbf{12.5}/13.0 & \multicolumn{1}{c|}{14.1/14.3} & \textbf{10.1}/10.3 \\ \hline \hline
\end{tabular}}
\vspace{-3mm}
\end{table*}

\subsection{System Configurations}
\label{subsec:systemconfigs}
\vspace{-.3em}

We considered two different geometry agnostic T-F mask estimation models as the baseline spectral masking systems: one predicting real-valued masks using sigmoid activation and the other estimating complex-valued masks. 
We used the mask estimation model from our recent paper \cite{yoshioka2021vararray}.
The model consists of nine consecutive conformer layers that were interleaved by two TAC layers, followed by mean pooling and another stack of six conformer layers. Each conformer layer used four heads with 64 dimensions and 33 convolution kernels. See \cite{yoshioka2021vararray} for further details\footnote{Note that our model setting is different from the one used in \cite{yoshioka2021vararray}, and thus the numbers cannot be directly compared.}.

As regards the ADL-MVDR component, 
we used 200 and 100 units in the two recurrent layers of $\mathbf{GRU{\text -}Net}_{\upsilon}$, followed by another 14-unit FFL with linear activation. For the $\mathbf{GRU{\text -}Net}_{\text{VV}}$, there were 200 units in both GRU layers with another 98-unit linear FFL. For VAD weights estimation, the corresponding GRU-Net featured a two-layer GRU with 200 units each, followed by a 1-unit (i.e., frequency-independent) FFL with ReLU activation. 

Two loss functions were examined; namely, the mean squared error (MSE) of the magnitude spectra~\cite{yu2017permutation} and the MSE of the log-mel features~\cite{Boeddeker18} obtained with 80 mel filter banks. 
We did not use the scale-invariant signal-to-distortion ratio (Si-SDR) loss which was used before for target speech extraction~\cite{zhang2021adl,zhang2020multi}. This is because the Si-SDR loss was found to be unstable when one source was silent as with the CSS task. $\alpha$ was set to 0.5 for residual connection. The AdamW optimizer \cite{loshchilov2017decoupled} was used with a weight decay of $1e^{-2}$. A warm-up learning schedule was used with the peak learning rate of 
$1e^{-3}$ followed by an exponential decay. 
The baseline systems were trained for 50 epochs, and the ADL-MVDR systems were trained jointly with the baseline systems for another 100 epochs. 
For CSS chunk-wise processing, $N_h,N_c,N_f$ were set to 1.2 s, 0.8s and 0.4 s, respectively.



\vspace{-.5em}
\section{Experimental Results}
\label{sec:experimentalresults}

\subsection{Results on LibriCSS}
\label{subsec:resultscss}
\vspace{-.3em}

The experimental results with different system configurations are shown in Table \ref{tab:tab1}.
Among the systems implementing some or all of the proposed improvements, the best performing approach was system 10 with a real-valued mask estimator. The proposed system achieved an approximately 9\% relative WER improvement on average (i.e., 10.1 \% vs. 11.1\%) compared with the baseline system (i.e., system 0 with real-valued mask). The proposed end-to-end neural beamforming was more advantageous for more challenging conditions, yielding a 15\% relative gain for the 30\% overlap condition (i.e., 12.5\% vs. 14.7\%). 

It is noteworthy that system 3 significantly outperformed system 2 (11.3 \% vs. 13.7\% for real-valued masks), and the latter even underperformed the baseline system. 
This indicates that the beamformer cannot fully eliminate the speech signals even when the output T-F masks are almost zero, especially in reverberant conditions \cite{yoshioka2018multi}. The use of the  VAD network helped the neural beamformer deal with the case of the output source being occasionally zero, as in the CSS task.

The results of Table \ref{tab:tab1} also show the impact of the normalized steering vector and applying the positive semi-definite constraint. Normalizing the estimated steering vector improved the ASR accuracy in most conditions. In particular, if we compare systems 3 and 4, we can see 3.5\% (i.e., 11.3\% vs. 10.9\%) and 3.6\% (i.e., 11.1\% vs. 10.7\%) relative WER gains for the real- and complex-valued mask configurations, respectively. Similar trends were observed for systems 7 and 8. 
Also, the positive semi-definite constraint on the estimate of $\mathbf{\Phi}^{-1}_{\mathrm{VV}}$ was found to preserve the performance with the reduced number of parameters. 

For the proposed end-to-end neural beamformers, the log-mel scale magnitude loss consistently yielded better ASR accuracy. For instance, between systems 10 and 6, there were 5.6\% and 6.4\% relative improvements for the real- and complex-valued mask configurations, respectively.

\subsection{Results on Real Meeting Recordings}
\label{subsec:resultsreal}
\vspace{-.3em}

\begin{table}[]
\centering
\caption{WERs (\%) on real meeting recordings.}
\vspace{-.7em}
\label{tab:tab2}
{\footnotesize
\begin{tabular}{l||c|c|c}
\hline \hline
System & MS & AMI-dev & AMI-eval \\ \hline
BF  &  17.8 & 25.0 & 27.3 \\
CSS w/ MVDR & 16.6 & \textbf{17.9} &  21.1 \\
CSS w/ ADL-MVDR & \textbf{16.3} & 18.1 & \textbf{20.6}  \\ \hline \hline
\end{tabular}
}
\vspace{-3mm}
\end{table}

For the real meeting experiments, we used a larger training set consisting of 438 hours of mixtures. 
Table \ref{tab:tab2} compares the proposed all-neural beamformer (i.e., system 10 with real-valued masks) with two baseline systems: one used the conventional chunk-wise MVDR as with \cite{yoshioka2021vararray}.
We also show the result with a super-directive beamformer with real-time beam steering, which only removes ambient noise and does not perform speech separation (denoted as BF).

For the MS dataset, the proposed ADL-MVDR achieved the best ASR accuracy (i.e., 16.3\%), which is about 8.4\% and 1.8\% relative improvements over BF and the conventional MVDR systems. While comparable performance was observed on the AMI development set for two MVDR systems, the proposed ADL-MVDR slightly outperformed the conventional MVDR by relative 2.4\% (i.e., 20.6\% vs. 21.1\%) in the AMI evaluation set. These results here suggest that the proposed neural beamformer achieved comparable ASR accuracy with a state-of-the-art MVDR-based system while also enabling frame-wise beamforming.

\section{Conclusions}
\label{sec:conclusions}
We developed the all-neural beamformer that enables frame-wise beamforming for the CSS task. Our neural beamformer achieved significant improvements in the ASR accuracy over baseline spectral masking systems, especially in challenging overlap conditions. For real conversation recordings, the proposed system achieved comparable performance to a conventional MVDR-based system with simplified runtime implementation. The experimental results suggest that VAD is needed for the neural beamformer to work effectively for CSS and that applying additional enhancements such as normalizing the steering vector further improved the performance. 

\bibliographystyle{IEEEtran}
{\small\bibliography{refs}}

\end{document}